\documentstyle[prb,aps,epsf,floats,goodfloat]{revtex}

\newcommand{\bc}{\begin{center}}
\newcommand{\ec}{\end{center}}
\newcommand{\be}{\begin{equation}}
\newcommand{\ee}{\end{equation}}
\newcommand{\beqn}{\begin{eqnarray}}
\newcommand{\eeqn}{\end{eqnarray}}
\begin{document}
\draft

\twocolumn[\hsize\textwidth\columnwidth\hsize\csname@twocolumnfalse%
\endcsname

\title{
Finite Temperature Ordering in the Three-Dimensional Gauge Glass 
}

\author{T. Olson and A. P. Young}
\address{Department of Physics, University of California, Santa Cruz, 
CA 95064}

\date{\today}

\maketitle

\begin{abstract}
    We present results of Monte Carlo simulations of the gauge glass model in
    three dimensions using exchange Monte Carlo.  We show for the first time
    clear evidence of the vortex glass ordered phase at finite temperature.
    Using finite size scaling we obtain estimates for the correlation length
    exponent, $\nu = 1.39 \pm 0.20$, the correlation function exponent,
    $\eta = -0.47 \pm 0.07 $, and  the dynamic exponent $z = 4.2 \pm
    0.6$.  Using our values for $z$ and $\nu$ we calculate the
    resistivity exponent to be $s = 4.5 \pm 1.1$.
    Finally, we provide a plausible lower
    bound on the the zero-temperature stiffness exponent, $\theta \ge 0.18$. 
\end{abstract}

\vskip 0.3 truein
]

\section{Introduction}

High-temperature superconductors have a phase diagram that is rich with 
physically diverse phenomena\cite{blatter}. 
In the mixed state of a pure 
type-II system one finds the Abrikosov lattice\cite{abrikosov}
of triangularly arranged 
vortices. This vortex lattice prohibits superconductivity because any 
perpendicularly applied current produces a Lorentz force causing the
vortices to move, dissipating energy. The addition of disorder, however,
drastically changes the behavior of the mixed state. Correlated disorder,
such as from heavy ion irradiation\cite{crabtree} or twin boundaries,\cite{grigera} 
causes the vortex lines
to locally align with and adhere to the defects; this destroys the long-range
order of the lattice and produces a superconducting glassy phase known 
as the Bose glass\cite{nelson}. Random point disorder, e.g., 
from proton irradiation
\cite{petrean}, pins the vortices in random positions creating a different
type of superconducting phase known as the vortex glass\cite{fisher,ffh}.

Similar to a spin glass, the ordered state of the vortex glass is characterized
by the phase of the superconducting order parameter randomly oriented in space
but frozen in time. A system must have a Hamiltonian that contains both 
randomness and frustration in order to exhibit this type of behavior. One such
model that has been used extensively to simulate the vortex glass transition is
the gauge glass\cite{blatter,huse_seung}. 

The gauge glass is important to study because like other spin glasses 
in 3 spatial dimensions,\cite{bhatt,kawa_young,mar_glass}
the existence of a finite temperature transition within the gauge glass is
controversial. The first estimate of the critical temperature was given in
1990 by Huse and Seung\cite{huse_seung}, $0 < T_c \le 0.6$. A year later this
estimate was improved significantly by Reger et al.,\cite{reger} who found that 
$T_c = 0.45 \pm 0.05$, but their data was insufficient to establish that spin
glass order occurs below this temperature.
Subsequent studies
\cite{gingras,cieplak,moore,kost_sim,maucourt,kos_ak} have focused
on defect energy scaling to
determine the existence of a finite $T_c$ by calculating the stiffness exponent
$\theta$. Unfortunately, these results seem to be dependent on the chosen 
boundary conditions, and there is little agreement on the values of $\theta$.
Consequently it is not clear whether the lower critical dimension 
is 3 or less than 3. 

In this paper we attempt to settle the controversy by studying the gauge glass 
using exchange Monte Carlo,\cite{huk_nemoto,hukushima} 
also known as parallel tempering.\cite{mar_et_al,marinari,newman} 
This technique allows us to
simulate lower temperatures and larger sizes than previously possible. The main
features of our work are as follows.
\begin{enumerate}
\item
We present for the first time clear evidence 
of vortex-glass ordering at finite temperature in the three-dimensional 
gauge glass model. We present superior data to that previously 
available\cite{reger} and obtain a more accurate determination of the
critical temperature $T_c/J = 0.47 \pm 0.03$.
\item
Using finite size scaling we determine the correlation length exponent
$\nu = 1.39 \pm 0.20 $ to a higher degree of precision then 
earlier works.\cite{wengel97,reger} The correlation 
function exponent is also obtained, $\eta = -0.47 \pm 0.07$. 
To our knowledge, this
is the first numerical estimate of $\eta$ for the gauge glass from simulations. 
Assuming hyperscaling, these two exponents completely determine the
universality class of the gauge glass model.
\item
Using standard Monte Carlo, we determine the dynamical exponent 
$z = 4.2 \pm 0.6$ 
and compare our results to experimental measurements
of the resistivity exponent $s = \nu (z - 1)$. We find $s = 4.5 \pm 1.1$.
\end{enumerate}

The layout of the paper is as follows. In \S\ref{sec-model}
we describe the model while in \S\ref{observable} we discuss the observables
that we measure.  In \S\ref{sec-exchange} 
we discuss our implementation of the exchange Monte Carlo
method and our tests for
equilibration. Our results for statics are discussed in 
\S\ref{sec-equilib} while our results for dynamics are given in
\S\ref{sec-dynamics}. In \S\ref{sec-concl} we summarize our results and give
some perspectives for future work.

\section{The Model}
\label{sec-model}

The gauge glass describes the physics of a disordered type-II
superconductor at distances larger than some characteristic length scale
$\Lambda$, beyond which order in the flux lattice has
been broken.\cite{blatter,larkin} One can then imagine the system as a granular 
superconductor (in an applied magnetic field) with an inter-grain 
separation of order $\Lambda$. Such a system can be modeled
as an array of nearest neighbor coupled Josephson junctions.
\cite{huse_seung,shih}
This leads to the Hamiltonian
\begin{equation}
\label{ham}
{\cal H} = -J \sum_{\langle i, j\rangle} \cos(\phi_i - \phi_j - A_{ij}),
\end{equation}
where each site $i$ on an $N = L\times L\times L$ cubic lattice has an 
associated phase angle $\phi_i$, $J$ is a positive ferromagnetic (Josephson) 
coupling between nearest neighbors, and $A_{ij}$ is proportional to the line
integral of the vector potential along 
a straight line path from site $i$ to its nearest neighbor site $j$,
\begin{equation}
A_{ij} = \frac{2 \pi}{\Phi_0} \int_{\vec{r}_i}^{\vec{r}_j} \vec{A} 
\cdot \vec{dl}.
\end{equation}
$\Phi_0 = h c/(2 e)$ is the flux quantum. 

The Hamiltonian of the gauge glass is given by Eq.~(\ref{ham})
in which the $A_{ij}$ {\em are quenched 
random variables} uniformly distributed from $0$ to $2 \pi$. Note that, by
contrast,
restricting the $A_{ij}$ only to values
$0$ and $\pi$ leads to the $XY$ spin glass, which is equivalent to setting
$A_{ij} = 0$
and taking the interactions to be $\pm J$ at random. 

The gauge glass is perhaps the simplest model of a disordered type-II
superconductor that contains the correct order parameter symmetry as well as
the randomness and frustration necessary for glassy behavior. However,
there are some features it ignores.\cite{huse_seung}

The model ignores screening since the $A_{ij}$ in
Eq.~(\ref{ham}) are quenched; there are no thermal fluctuations in the 
magnetic field. This corresponds to the extreme type-II limit in which the 
Ginzburg-Landau\cite{ginzburg}  parameter
$\kappa = \lambda/\xi\rightarrow \infty$, where $\lambda$ 
is the penetration depth and $\xi$ is the coherence length. This limit
may be realistic since $\kappa$ can be quite large in high $T_c$
superconductors, e.g., $\kappa \approx 90$ for 
$\mathrm{YBa_2Cu_3O_{7-\delta}}$. It seems, however, that when the
gauge glass is modified to include strong screening, 
the finite temperature transition to the vortex glass 
phase is rounded out in three dimensions very close to the putative
$T_c$.\cite{by,wengel,wengel97,kisker,pfeiffer,kawamura} This rounding 
probably takes place over such a narrow temperature region that will be 
very difficult to observe. Hence a model which neglects screening, like
the gauge glass, should be
able to account for most of the observable data in the critical region.

Unlike a real superconductor in a magnetic field, the gauge glass is 
isotropic on average. There are local quenched fluxes but no net field in any
direction.
In six or more spatial dimensions the lack of anisotropy does not
seem to matter\cite{john_lub}; however it is still an open question 
whether, in three dimensions,
the vortex glass transition in the gauge glass
and in a system with a non-zero net field
are in the same universality class.

The source of the quenched randomness in the gauge glass model is the 
vector potentials linking the sites. This is not very realistic, and a more
accurate model would have vector potentials corresponding to a uniform
field, and put disorder into the strength of the couplings. However, the
precise
details of the disorder should be irrelevant 
for critical phenomena

In this paper we shall show very clearly that the gauge glass has a 
finite temperature transition to a vortex glass ordered state and that its 
critical exponents agree, within fairly large error bars,
with some experimental measurements on 
type-II superconductors. More work remains to be done to check whether 
an anisotropic model with a net field would change the
universality class.

\section{The Observables}
\label{observable}

A standard technique to determine the critical 
temperature is to use the Binder ratio\cite{binder}
\begin{equation}
g(L) = 2 - \frac{\lbrack \langle \left|q\right|^4 \rangle 
\rbrack_{\mathrm{av}}}{\lbrack \langle \left|q\right|^2 \rangle 
\rbrack_{\mathrm{av}}^2} ,
\end{equation}
where $\lbrack \cdots \rbrack_{\mathrm{av}}$ denotes an average over 
configurations of the disorder, $\langle \cdots \rangle$ denotes a thermal 
average and $q$ is the complex overlap order parameter
\begin{equation}
q = \frac{1}{N}\sum_{j=1}^N \exp[i(\phi_j^\alpha - \phi_j^\beta)] ,
\label{q_def}
\end{equation}
in which $\alpha$ and $\beta$ denote two independent replicas with the same
disorder.
One plots $g(L)$ vs. $T$ for
different $L$ and identifies the temperature at which the curves cross as
$T_c$. However, since $g(L)$ cannot exceed unity,
the splaying out of the data for $g(L)$
below $T_c$ (which indicates spin glass order) is a small effect which can be
difficult to see. This is why it as been so hard to establish conclusively that
there is spin glass order at finite-$T$ in the
three-dimensional Ising 
spin-glass.\cite{bhatt,kawa_young,mar_glass}

In order to avoid this problem we
follow Reger et al.\cite{reger} in calculating a current; this is defined as
the rate of change of the free
energy with respect to a twist angle $\Theta$ at 
the boundaries. We begin
by replacing periodic boundary conditions with twisted 
boundary conditions along one of the axes $\hat{x}$, i.e.
\begin{equation}
\phi_{i+L\hat{x}} = \phi_i +  \Theta. 
\end{equation}
Note that $\Theta = 0$ corresponds to periodic boundary conditions, and
$\Theta = \pi$ corresponds to anti-periodic boundary conditions.
We can convert the model back to periodic boundary conditions by redefining 
$\phi$ through
\begin{equation}
\phi_{i+n\hat{x}} \rightarrow \phi_{i+n\hat{x}} - {n \over L} \Theta.\\
\end{equation}
The model then is precisely Eq.~(\ref{ham})
with periodic boundary conditions but with
the $A_{ij}$ for bonds in the $x$-direction changed according to
\begin{equation}
A_{i,i+\hat{x}} \rightarrow A_{i,i+\hat{x}} - {\Theta \over L }.
\end{equation}
Using $F = -\ln(Z)/\beta$ we define a current as the response of the free energy
to an infinitesimal $\Theta$,
\begin{equation}
I(L) \equiv \lim_{\Theta \rightarrow 0} \frac{\partial F}{\partial \Theta}=
\frac{1}{L}\sum_i \langle \sin (\phi_i-\phi_{i+\hat{x}}-A_{i,i+\hat{x}})\rangle.
\end{equation}

Because the $A_{ij}$
are uniformly distributed over the entire period of the sine function, the
the value of the current averaged over samples is zero, i.e.
\begin{equation}
\lbrack I(L) \rbrack_{\mathrm{av}} = 0.
\end{equation}
Consequently we calculate the root-mean-square current, given by
\begin{equation}
\label{i_rms}
I_{\mathrm{rms}} \equiv \sqrt{\lbrack \langle I_\alpha(L) \rangle
\langle I_\beta(L) \rangle \rbrack_{\mathrm{av}}} ,
\end{equation}
where $\alpha$ and $\beta$ denote two independent replicas with the same
disorder. We use two replicas to avoid any bias in calculating the average
of the square of the current.

The primary advantage of using $I_{\mathrm{rms}}$ rather than $g(L)$ is that
$I_{\mathrm{rms}}$ increases with $L$ for $T < T_c$, so the splaying out the
data below $T_c$ should  be much easier to see than for the Binder ratio.
In the ordered phase the current should scale
with the stiffness
exponent\cite{fh,gingras,cieplak,moore,kost_sim,maucourt,kos_ak}
$\theta$, i.e.
\begin{equation}
\label{i_lowT}
I_{\mathrm{rms}} \sim L^\theta, \qquad \qquad (T < T_c) ,
\end{equation}
where $\theta > 0$ if $T_c > 0$.
Above $T_c$, where spin-spin correlations are small,
we expect $I_{\mathrm{rms}}$ to approach zero with increasing $L$
because larger systems are less sensitive to a twist at the boundaries.
The $I_{\mathrm{rms}}$ curves cross at $T_c$ and have the finite size
scaling form
\begin{equation}
\label{fss_eqn}
I_{\mathrm{rms}} = \tilde{I}(L^{1/\nu} t)
\end{equation}
where
\begin{equation}
t = (T - T_c)/T_c  
\end{equation}
is the reduced temperature.

\section{Exchange Monte Carlo and Equilibration}
\label{sec-exchange}

Exchange Monte Carlo,\cite{huk_nemoto,hukushima} also called parallel 
tempering,\cite{mar_et_al,marinari,newman} is a technique 
for simultaneously simulating multiple copies of a particular 
configuration of disorder with each copy at a different temperature.
After a certain number of sweeps through the lattice, one tries to exchange
the spin configurations of copies at neighboring temperatures with the
probability
\begin{equation}
P(\sigma_m \leftrightarrow \sigma_{m+1};\beta_m,\beta_{m+1}) = 
\exp(-\Delta)
\end{equation}
where
\begin{equation}
\Delta \equiv (\beta_{m}-\beta_{m+1})(E(\sigma_{m+1})- E(\sigma_m)).
\end{equation}
$\sigma_m$ is the spin configuration of the $m$th copy which has inverse
temperature
$\beta_m$ and total energy $E(\sigma_m)$.
A given spin configuration is thus heated and cooled many times during the
simulation. Since equilibration is fast at high temperature, each time
the system is cooled the minimum (valley) that it visits is uncorrelated
with the minimum that it visited at the previous cooling.
Hence the system can can visit
different local minima at low temperature more efficiently than if the
temperature were kept fixed. In the latter case, very large barriers would have
to be overcome, which takes a time {\em exponentially}
large in the ratio of the
barrier height to the temperature.

In deciding what temperatures to simulate, one would like
the energy distributions of neighboring temperatures to have
enough overlap that the probability to exchange configurations
is sufficiently high. This requires

\begin{equation}
{T_{m+1} - T_m \over T_m } = c_m N^{-1/2},
\end{equation}
where the $c_m$ are constants of order unity. In our simulations we took all
the $c_m$ to be exactly unity, which gave a satisfactory
acceptance ratio for temperature exchanges 
in the interval from $0.5$ to $0.6$ for all sizes and temperatures studied.

We took the highest temperature, $T_{\mathrm{max}}$, to be approximately $2
T_c$ at which the spin dynamics are quite rapid. Temperature
exchanges were carried out after every 10 Monte Carlo sweeps, after
which ``time"
the normalized energy-energy auto-correlation function at $T_{\mathrm{max}}$ 
is quite low, about 0.25.

\begin{figure}
\epsfxsize=\columnwidth\epsfbox{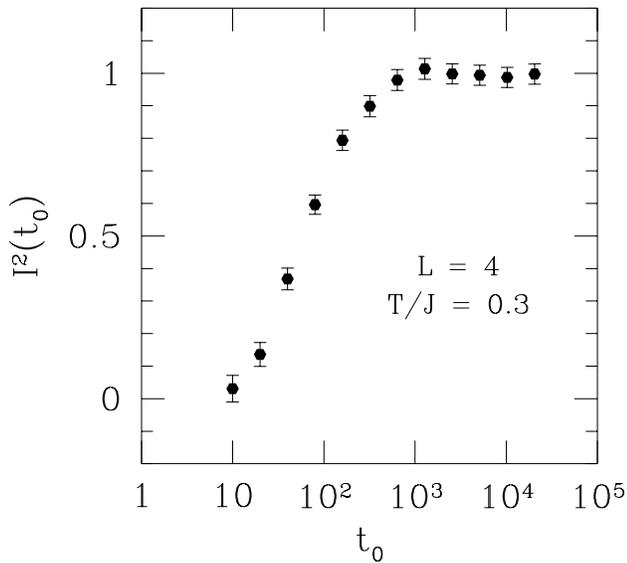}
\caption{
A semi-log plot of $I^2(t_0)$ against $t_0$,
from Eq.~(\ref{i_t0}) for $L = 4$ at $T/J = 0.3$ 
using $2000$ samples to test for equilibration.
One sees that the results do not change for $t_0$ greater than about 1000.
Hence, in the production runs, the equilibration time was taken to
be $t_0 = 1280$.
}
\label{i_equilib}
\end{figure}

The equilibration time, $t_{\mathrm{eq}}$, for $I_{\mathrm{rms}}$ 
is determined by following the 
temporal evolution of 
\begin{equation}
\label{i_t0}
I^2(t_0) \equiv \left[ \frac{1}{t_0} \sum_{t=1}^{t_0} 
I_\alpha(t+t_0) I_\beta(t+t_0) 
\right]_{\mathrm{av}},
\end{equation} 
where $t_0$ is the number of equilibration sweeps as well the number 
of measurement sweeps and the subscripts $\alpha$ and $\beta$ denote
independent replicas. When $t_0 \ll t_{\mathrm{eq}}$ 
the spin configurations of the two replicas are completely uncorrelated
but as $t_0$ increases 
they become more correlated, since they both feel the same random interactions.
Thus we expect $I^2(t_0)$ to monotonically increase from zero to the equilibrium
value as $t_0 \rightarrow t_{\mathrm{eq}}$; see Fig.~(\ref{i_equilib}).
Each of the equilibration times in Table~\ref{pt_temps} is chosen to be the
least number of sweeps necessary to equilibrate at the lowest
temperature, $T_{\mathrm{min}}$. For each size except $L = 12$
we ran some samples for $t_0 \gg  t_{\mathrm{eq}}$ in order to confidently
determine $t_{\mathrm{eq}}$;
the remaining samples were run with $t_0 = t_{\mathrm{eq}}$.
We should note that for $L = 12$ it was impractical to run
jobs for $t_0$ much greater than the value at which the $I^2(t_0)$ seemed to
stop increasing, so we took this value to be $t_{\mathrm{eq}}$.

\begin{table}
\begin{tabular}{|c|c|c|c|c|c|}
$L$ & $T_{\mathrm{min}}$ & $T_{\mathrm{max}}$ & $n_T$ & $t_{\mathrm{eq}}$ & Samples \\
\hline
4 & 0.3 & 0.97 & 11 & 1280 & 8000 \\ 
6 & 0.3 & 0.92 & 18 & 10240 & 10937 \\
8 & 0.3 & 0.92 & 27 & 20480 & 5388 \\
12 & 0.3 & 0.90 & 47 & 163840 & 781 \\
\end{tabular}
\caption
{
Parameters of
the exchange Monte Carlo simulations for each value of $L$, 
where $T_{\mathrm{min}}$ and $T_{\mathrm{max}}$ are the minimum and
maximum of $n_T$ temperatures,
and $t_{\mathrm{eq}}$ is the number of sweeps for equilibration, which, in our
simulations, is also
equal to the number of sweeps for measurements. The number of samples studied
for each size is also shown.
\label{pt_temps}
}
\end{table}

\section{Results for Statics}
\label{sec-equilib}

\begin{figure}
\epsfxsize=\columnwidth\epsfbox{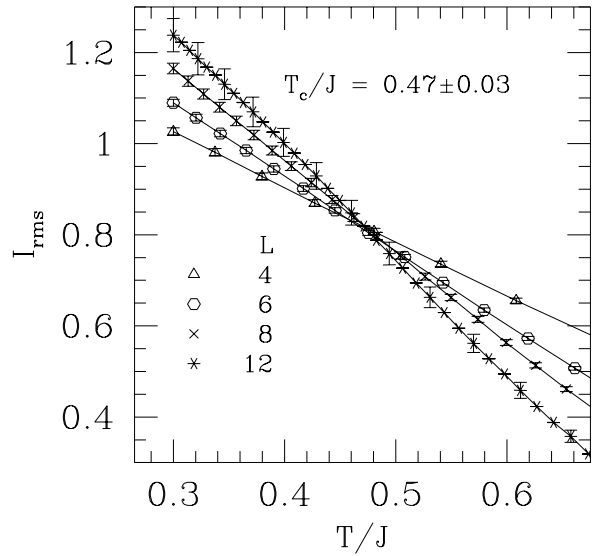}
\caption{
A plot of $I_{\mathrm{rms}}$ from Eq.~(\ref{i_rms})
for $L = 4, 6, 8, \mbox{and }12$.
For the point of intersection we estimate that
the critical temperature is $0.47 \pm 0.03$. Some of the $L = 12$ errorbars
have been removed for clarity.
}
\label{currents}
\end{figure}

We present our results for $I_{\mathrm{rms}}$ from 
Eq.~(\ref{i_rms}) in Fig.~(\ref{currents}).
The data cross at $T_c/J = 0.47$ and
splay out clearly on {\em both}
sides of the transition. This is the first time that
the splaying out of the data for $I_{\mathrm{rms}}$
in the ordered state can be distinguished well beyond
the size of the error bars, thus presenting incontrovertible 
evidence for a spin-glass ordered phase in three dimensions. The data for
the lowest temperature
are given numerically in Table~(\ref{lowT}).
We obtain an uncertainty of $0.03$ 
for $T_c/J$ by setting the error equal to the region over which 
the data for all sizes overlap within their error bars. 
Our final estimate for the critical temperature is
\begin{equation}
\label{Tc_eqn}
T_c/J = 0.47 \pm 0.03,
\end{equation}
which agrees with the previous value of $0.45 \pm 0.05$ from 
Reger et al.\cite{reger} The key difference is that in Reger et al.
the values for $I_{\mathrm{rms}}$ did not splay out significantly below $T_c$
and so they did not find compelling evidence for spin glass order.

\begin{table}
\begin{tabular}{|c|c|c|c|}
$L = 4$ & $L = 6$ & $L = 8$ & $L = 12$ \\
\hline
$1.026 \pm 0.009$ & $1.090 \pm 0.008$ & $1.165 \pm 0.012$ 
& $1.238 \pm 0.037$ \\ 
\end{tabular}
\caption
{
$I_{\mathrm{rms}}$ from  Eq.~(\ref{i_rms}) for $T/J = 0.3$ and different $L$. 
The values are clearly distinct beyond the size of the error-bars. 
\label{lowT}
}
\end{table}

\begin{figure}
\epsfxsize=\columnwidth\epsfbox{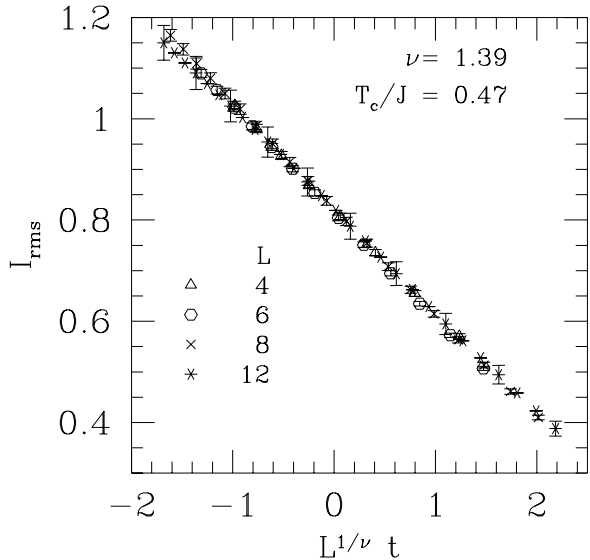}
\caption{
A plot of $I_{\mathrm{rms}}$ from Eq.~(\ref{i_rms}) versus $L^{1/\nu} t$, where
$t = (T - T_c)/T_c$ and $\nu = 1.39$.
Some of the $L = 12$ errorbars
have been removed for clarity.
}
\label{i_fss}
\end{figure}

We obtain an estimate for the correlation length exponent $\nu$ by finite-size
scaling the data from Fig.~(\ref{currents}); see also Eq.~(\ref{fss_eqn}). 
Fig.~(\ref{i_fss}) shows that the data scales well with $\nu = 1.39$. This
value is
obtained by by scaling the data using different choices
for $\nu$, fitting a polynomial, calculating 
the chi-squared statistic for each choice, and minimizing the
chi-squared. The error in $\nu$ is determined by
varying the value of
$\nu$ until the $L=8$ data no longer scales with the smaller
sizes within the sum of their errorbars. $L=8$ was chosen rather than
$L=12$ because the errorbars of the former are much smaller. This leads to our
estimate
\begin{equation}
\label{nu_eqn}
\nu = 1.39 \pm 0.20 .
\end{equation} 
Variations in
$T_c$ from Eq.~(\ref{Tc_eqn}) did not significantly
increase the error bar beyond what that shown in Eq.~(\ref{nu_eqn}).
Our result agrees with, and is a bit more precise than, those previously 
presented by Wengel and Young,\cite{wengel97} $\nu = 1.3 \pm 0.3$, and by
Reger et al,\cite{reger} $ 1.3 \pm 0.4$.

The correlation length exponent has also
been deduced from experimental measurements.
It is now known,\cite{crabtree,grigera} however, that the
response of the resistivity and critical temperature to tilting of the applied
field distinguishes a vortex-glass transition, in which point disorder
dominates, from a Bose-glass transition, in which correlated disorder, such as
twin boundaries or columnar pins, dominates. 
To our knowledge, there are only two experiments\cite{petrean,klein} 
which observe a vortex glass transition {\em and} demonstrate the proper 
response to magnetic
field tilting.

In the first, Petrean et al.\cite{petrean} find that for
untwinned proton-irradiated $\mathrm{YBa_2Cu_3O_{7-\delta}}$ 
near criticality the resistivity decreases from its maximum 
as the field is tilted away from the c axis,\cite{petrean},
which signals a vortex
glass transition, whereas in twinned
$\mathrm{YBa_2Cu_3O_{7-\delta}}$, they find
a Bose-glass transition with the resistivity
increasing from its minimum.\cite{grigera}
Petrean et al.,\cite{petrean} do not obtain $\nu$ but do obtain the resistivity
exponent $s$ by fitting resistivity vs. temperature curves. We will compare our
results to these in \S\ref{sec-dynamics}.  

In the second work, Klein et al.,\cite{klein} study
the vortex glass transition in $\mathrm{(K,Ba)BiO_3}$. The dependence of their
data on field tilting is that expected for the vortex glass, and they obtain
$\nu = 1.0 \pm 0.2$ which
agrees with ours within the errorbars.

Kawamura\cite{kawamura} has recently modified the gauge glass to include a net 
magnetic field. He obtains $\nu = 2.2 \pm 0.4$ which is greater
than our estimate and the experimental result of Klein et al\cite{klein}.   

\begin{figure}
\epsfxsize=\columnwidth\epsfbox{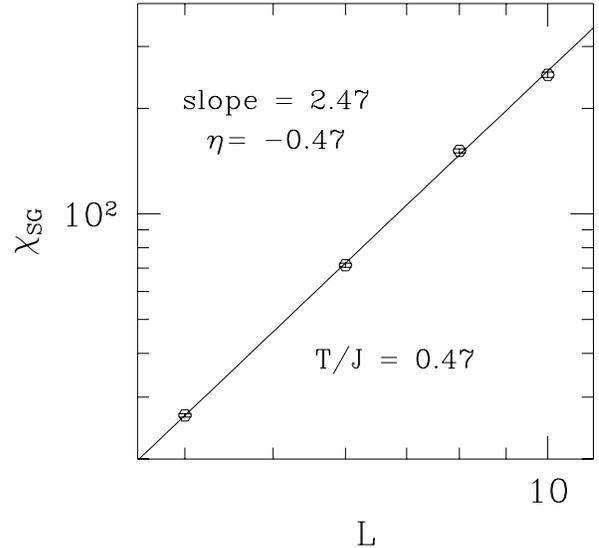}
\caption{
A log-log plot of $\chi_{\mathrm{SG}}$
vs. $L$ for $L = 4,6,8, \mbox{and } 10$ at
$T/J = 0.47$
according to Eq.~(\ref{suscept_eqn})
with $\eta = -0.47$. 
}
\label{eta_plot}
\end{figure}

In addition to $\nu$, we calculate the independent exponent $\eta$ which 
describes the decay of the correlation function at criticality.
To our knowledge,
this is the first calculation of $\eta$ for the gauge glass in three
dimensions.
Finite size scaling predicts that, at the critical point, the
spin glass susceptibility 
\begin{equation}
\label{sg_sus}
\chi_{\mathrm{SG}} = N \left[ \langle |q|^2 \rangle \right]_{\mathrm{av}}
\end{equation}
scales as
\begin{equation}
\label{suscept_eqn}
\chi_{\mathrm{SG}} \sim L^{2 - \eta},
\end{equation}
assuming the hyperscaling relation $\gamma/\nu = 2 - \eta$. We use
standard Monte Carlo to calculate the susceptibility at $T/J = 0.44, 0.47,
\mbox{and } 0.50$;
the details
of these simulations are discussed in \S\ref{sec-dynamics}.
We present our results in Fig.~(\ref{eta_plot}). From linear least-squares fits
to the
data on a log-log scale
we obtain
\begin{equation}
\label{eta}
\eta = -0.47 \pm 0.07.
\end{equation}
The error in our estimate comes mainly from the uncertainty in $T_c$.
Kawamura's\cite{kawamura} anisotropic gauge glass yields a
similar estimate, $\eta = -0.5 \pm 0.2$.

There has been some controversy regarding the value of the zero temperature
stiffness exponent $\theta$. This is typically computed from the
root-mean-square ``defect energy'' on changing the boundary conditions from
periodic to anti-periodic, but is also given by the response to an
infinitesimal twist in the boundary conditions, as shown in Eq.~(\ref{i_lowT}).
A positive value for $\theta$ indicates a finite temperature transition to
a spin glass state, whereas $\theta < 0$ implies $T_c = 0$.
Some studies\cite{reger,gingras,kost_sim,maucourt}
obtain $0 \le \theta \le0.077$; we refer to these as the low group. Others,
\cite{cieplak,moore,kos_ak} however, calculate $0.26 \le \theta \le 0.31$; 
we name these the high group.

\begin{figure}
\epsfxsize=\columnwidth\epsfbox{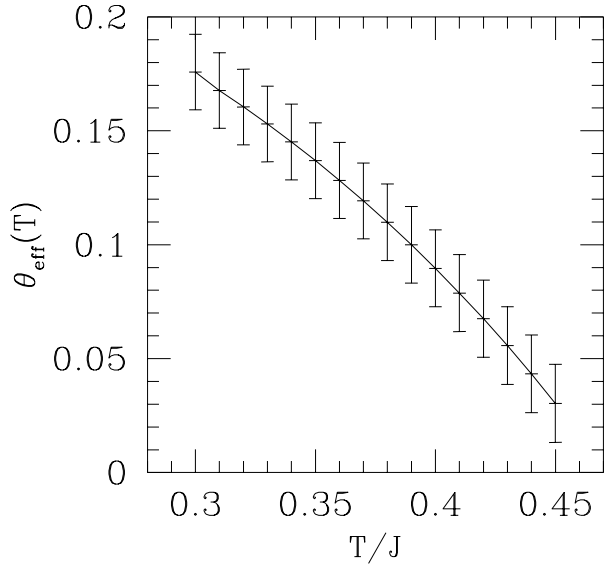}
\caption{
A plot of $\theta_{\mathrm{eff}}(T)$ vs. $T$. At each $T$, 
$\theta_{\mathrm{eff}}(T)$ is obtained from a linear least squares
fit of $\log I_{\mathrm{rms}}$ against $\log L$.
\label{theta_est}
}
\end{figure}

Our results for the rms current are not at
sufficiently low $T$ and large $L$
to get a firm estimate for $\theta$, but we obtain a plausible bound as
follows. For each temperature below $T_c$ we do a linear least squares fit of
$\log I_{\mathrm{rms}}$ against $\log L$ to get an {\em effective} temperature
dependent value, $\theta_{\mathrm{eff}}(T)$. The results are shown in 
Fig.~(\ref{theta_est}). We see that $\theta_{\mathrm{eff}}$ increases
monotonically as $T$ decreases, so we expect that the asymptotic value
is greater than the value of $\theta_{\mathrm{eff}}$ at the lowest temperature, i.e. we expect
\begin{equation}
\theta \ge 0.18 ,
\end{equation}
which is consistent with the results of the ``high group''.

\section{Dynamics}
\label{sec-dynamics}
When the gauge glass is sufficiently close to criticality the 
correlation length is of order $L$ and the system experiences critical slowing
down.  The equilibration time $t_{\mathrm{eq}}$ then scales as 
\begin{equation}
\label{slow_down}
t_{\mathrm{eq}} \sim L^z,
\end{equation}
where $z$ is the dynamic exponent. At temperatures just above the 
vortex glass transition, the resistivity $\rho$ is predicted\cite{ffh} 
to scale as
\begin{equation}
\rho \sim (T - T_c)^s,
\end{equation}
where $s$ is related to other exponents by
\begin{equation}
\label{s_eqn}
s = \nu (z - 1)
\end{equation}
in three dimensions.
We are thus motivated to calculate $z$ to compare our results 
to experimental measurements of $s$. It must be emphasized,
however, that there are more dynamical universality classes than static
universality classes. In our simulations which determine $z$ we use
dissipative (Monte Carlo) dynamics
with standard Metropolis-type updating probabilities, and {\em without} the
temperature swapping (exchange Monte Carlo) that we used in our
simulations of static quantities. Tests for equilibration were carried out as
in Bhatt and Young\cite{bhatt}.

\begin{table}
\begin{tabular}{|c||c|c|c|c|}
$T/J$ & $L = 4$ & $L = 6$ & $L = 8$ & $L = 10$ \\
\hline
$0.44$ & $500$ & $400$ & $358$ & $372$ \\ 
$0.47$ & $500$ & $400$ & $328$ & $246$ \\ 
$0.50$ & $500$ & $400$ & $350$ & $321$ \\ 
\end{tabular}
\caption
{
Number of samples in the standard Monte Carlo simulations used for
calculating the dynamic exponent. Three temperatures were used to 
account for the uncertainty in $T_c$.
\label{mc_stats}
}
\end{table}

The dynamic exponent can be obtained by a finite size scaling
of time-dependent
measurements of the spin glass susceptibility.
We use the ``two-replica'' susceptibility $\chi_{\mathrm{SG}}(t_0)$ defined by
\begin{equation}
\label{chi_over}
\chi_{\mathrm{SG}}(t_0) = \left[ \frac{1}{N t_0} \sum_{t=1}^{t_0}
\left| \sum_{j=1}^N e^{ i [\phi_j^\alpha(t_0+t) - \phi_j^\beta(t_0+t) ] } 
\right|^2 \right]_{\mathrm{av}}
\end{equation} 
where $\alpha$ and $\beta$ denote independent replicas.

\begin{figure}
\epsfxsize=\columnwidth\epsfbox{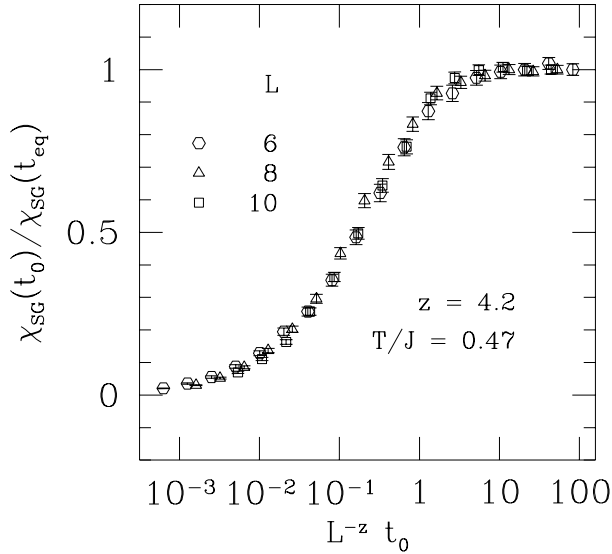}
\caption{
A finite-size scaling plot of $\chi_{\mathrm{SG}}(t_0)$ using 
Eq.~(\ref{chi_fss_eqn}) for $L = 6,8,\mbox{and }10$ at $T/J = 0.47$
with $z = 4.2$. 
}
\label{z_plot}
\end{figure}

We obtain our estimate of $z$ by a finite-size scaling analysis of the data for
$\chi_{\mathrm{SG}}(t_0)$ 
for sizes $L = 6,8,\mbox{and }10$;
the scaling is better without the $L = 4$
data, so it has been omitted. One expects
from Eq.~(\ref{slow_down}) that at criticality 
\begin{equation}
\label{chi_fss_eqn}
\frac{\chi_{\mathrm{SG}}(t_0)}{\chi_{\mathrm{SG}}(t_{\mathrm{eq}})} =
f(L^{-z} t_0)
\end{equation}
where $f$ is a scaling function. Our results, shown in Fig.~(\ref{z_plot}),
yield
\begin{equation}
\label{z_eqn}
z = 4.2 \pm  0.6,
\end{equation}
as calculated from the $T/J=0.47$ data. The estimate of the uncertainty
is obtained by varying $z$ until the $L = 10$ data no longer scales with
the smaller sizes within the errorbars. Similar scaling plots from the
$T/J = 0.44$ and $T/J = 0.50$ yield $z$ values within
Eq.~(\ref{z_eqn}), thus the uncertainty in $T_c$ is not the dominant
contribution to the error in $z$. 
Our value agrees with the previous estimate $z = 4.7 \pm 0.7$ 
by Reger et al.\cite{reger}
and $z = 3.3 \pm 0.5$ from Kawamura's\cite{kawamura} anisotropic gauge glass. 

Using Eqs.~(\ref{z_eqn}) and (\ref{nu_eqn}), the resistivity exponent 
is obtained from Eq.~(\ref{s_eqn}),
\begin{equation}
s = 4.5 \pm 1.1
\end{equation}
This value agrees with the experiments of Klein et al.\cite{klein},
$s = 3.9 \pm 0.3$, on $\mathrm{(K,Ba)BiO_3}$,
as well as Petrean et al.,\cite{petrean} $s = 5.3 \pm 0.7$, 
who study untwinned
proton-irradiated $\mathrm{YBa_2Cu_3O_{7-\delta}}$.
 
\section{Conclusions}
\label{sec-concl}
In this work we have presented exchange Monte Carlo results for the
gauge glass model in three dimensions. We have shown for the
first time incontrovertible evidence of a vortex-glass ordered
phase below $T_c$. To our knowledge, this is the first time
that such clear ordering has been shown for 
{\em any} spin-glass model in three dimensions.
The correlation length exponent $\nu$ has been 
calculated to higher precision than before. We have also obtained
the first estimate of the correlation function exponent $\eta$
from Monte Carlo simulations.
Finally, our values of $\nu$
and $z$ are combined to estimate the resistivity exponent $s$. Our results
are summarized in Table~(\ref{summary}). 

It is interesting that our values of $\eta$ and $\nu$ agree with those of
two other glassy systems, the three-dimensional $\pm J$ 
Ising spin glass\cite{kawa_young} and chiral ordering 
of the three-dimensional $\pm J XY$ spin glass.\cite{kawa_chiral} 
Kawashima and Young\cite{kawa_young} obtain $\nu = 1.7 \pm 0.3$ and
$\eta = -0.35 \pm 0.05$ for the Ising spin glass, while 
Kawamura\cite{kawa_chiral} obtains $\nu = 1.5 \pm 0.3$ and
$\eta = -0.4 \pm 0.2$ for the chiral glass ordering. However, it
is expected that
the Ising and gauge glass
models are in different universality classes because the order
parameter has a different number of components in each case: one for the Ising
spin glass, and two (the real and imaginary parts of Eq.~(\ref{q_def}))
for the gauge glass. 
Indeed, a first
order $\epsilon = 6 - d$ expansion shows that the gauge glass exponents
are not the same as any $n$-component vector spin glass.\cite{houghton}
The order parameters for the Ising spin glass and the chiral
transition in the XY spin glass do have the same dimensionality, but
the transitions are still likely
to be in different universality classes because of the long range interaction
between chiralities in the XY spin glass.
The error bars in the exponents are not extremely small so
the apparent agreement
between the results may be simply a numerical coincidence.
\begin{table}
\begin{tabular}{|c|c|c|c|}
$T_c/J$ & $\nu$ & $\eta$ & $z$ \\
\hline
$0.47 \pm 0.03$ & $1.39 \pm 0.20$ & $-0.47 \pm 0.07$ 
& $4.2 \pm 0.6$ \\ 
\end{tabular}
\caption
{
Critical temperature and exponents of the gauge glass in three dimensions.
\label{summary}
}
\end{table}

To confidently compare these results with experiment
one has to show that the
critical point in the gauge glass model is in the same universality class as an
anisotropic model with a net field. Clearly, if
such anisotropy causes the scaling behavior to be
anisotropic, in the sense that the {\em exponents} for the divergence of the
correlation length are different for separations along and perpendicular to the
field, then it is relevant.  If, however, the anisotropy induced by a net
field
does not lead to anisotropic scaling, but
just causes the {\em amplitudes} of the correlation lengths in the different
directions to be different, then it is not clear why it should be relevant,
just as making an Ising ferromagnet anisotropic by having the bonds in one
direction different from those in the other directions does not change the
universality class.

Kawamura\cite{kawamura} has studied a vortex glass model with a net field and
finds different critical behavior from that of the gauge glass, even though the
scaling is {\em not} anisotropic in the sense defined above.  However, he
imposes free boundary conditions which may lead to very large corrections to
finite-size scaling (since a large fraction of the sites are on the boundary).
Hence it would be very useful to study the vortex glass transition in a
model with a net field {\em and} periodic boundary conditions.

\acknowledgments
We would like to thank W.~K.~Kwok for useful discussions regarding experimental
data.  This work was supported by the National Science Foundation under grant
DMR 9713977.

\end{document}